\documentclass[aps,prl,twocolumn,showpacs,preprintnumbers,amsmath,amssymb]{revtex4-2}
\usepackage{graphicx}
\usepackage{dcolumn}
\usepackage{bm}

\begin{document}

\title{Apsidal motion and physical parameters in the eclipsing system
V490~Sct}%


\author{Igor M. Volkov}
\affiliation{Sternberg Astronomical Institute, Lomonosov Moscow State University,
Universitetskii pr.13, 119991 Moscow, Russia}

\affiliation{Institute of Astronomy of the Russian Academy of Sciences, 
Pyatnitskaya str.48, 119017 Moscow, Russia}

\author{Alexandra S. Kravtsova}
\affiliation{Sternberg Astronomical Institute,
Lomonosov Moscow State University,
Universitetskii pr.13, 119991 Moscow, Russia}

\affiliation{Institute of Astronomy of the Russian Academy of Sciences,
Pyatnitskaya str.48, 119017 Moscow, Russia}

\date{\today}

\begin{abstract}
We report long-termed $UBVRIRcIc$ photometry of the highly
eccentric 12.04 day detached eclipsing binary V490~Sct ($V$=13.1,
B9.5+A0, $e$ = 0.40), which we use to determine its relative and
absolute parameters. The absolute masses, radii, and temperatures
are $M_A$~=~2.33$\pm0.1$ M$_\odot$, $R_A$~=~1.91$\pm0.04$
$R_\odot$, and $T_A$~=~9960$\pm$60 K for the primary and
$M_B$~=~2.24$\pm0.1$ M$_\odot$, $R_B$~=~1.86$\pm0.04$ $R_\odot$,
and $T_B$~=~9700$\pm$80 K for the secondary. The system displays a
slow periastron advance that is dominated by general relativity
(GR). Our measurement, $\dot\omega~=~0.86$ deg century$^{-1}$, is
32\% less then the expected rate, $\dot\omega~=~1.24$ deg
century$^{-1}$, which has an 83\% contribution from GR. A
comparison with current stellar evolution models shows a good
match to the measured properties at an age of about 130~mln. years
and Solar abundance. The photometrical parallax of the system
$\pi=0.77\pm0.02$ mas, matches quite well the \textit{GAIA} DR2
value, $\pi=0.76\pm0.04$ mas.
\end{abstract}


\maketitle
\section{Introduction}
V490~Sct was found to be variable by Dr H. van Gent who has
investigated the variability of the stars in a region of 100
square degrees in the constellation of Sagittarius around the
central star BD$-18^\circ 5206$. Most of the plates (382 pieces)
were obtained by him with the help of the Franklin-Adams camera
(D=25cm) of the Union Observatory. 8 more plates were received by
P.~Th.~Oosterhoff with the Mount Wilson 10-inch refractor.
J.~Uiterdijk has investigated all new variables on these plates.
He found that the star with serial number 42 in his list was the
eclipsing variable. He derived its true period and due to
displacement of the secondary minimum estimated its eccentricity
as 0.4, \citet{1949AnLei..20...41U}. He noted that the data
obtained need confirmation so he published individual observations
in minima for further use. These relatively inaccurate
photographical estimates obtained by the Neiland-Blazko method
have played a role in our study of the apsidal motion due to the
fact that they are 80 years away from the epoch of our
observations. The star was firstly designated as V1049~Sgr but
when GCVS research group realized that it is situated in Scutum
constellation 8' west from the Sagittarius border it was renamed
to V490~Sct \citet{2017ARep...61...80S}. Based on Uiterdijk's
data, the star was included in the lists of promising for internal
structure and relativistic effect investigation objects such as
\citet{1994ExA.....5...91G} (named as V1049~Sgr),
\citet{2018ApJS..235...41K} (named as V490~Sct) and several other
catalogues. No further researches followed the work of Uiterdijk
although from the very beginning it was clear that due to the
significant eccentricity and favorable orientation of the orbital
ellipse, the star is a very promising object for the internal
structure studying and general relativity (GR) testing.

\section{Observations and data reduction}

We included the star in our program of eclipsing eccentrical
systems study, \citet{2009ARep...53..136V}. The observations
started as far as in 1989 year at Tien-Shan high altitude
observatory of SAI. That time we failed to detect minima. We
continued occasional observations of the star according the
Uiterdijk's ephemeris, but things did not get off the ground until
we began systematic observations of the star in 2004 in Crimean
observatory of SAI and in Simeiz observatory of INASAN regardless
the predictions of the ephemeris. In 2005 we finally found one of
the minima and a meaningful accumulation of the observational data
has begun. It turned out that the initial ephemeris gave an error
of 10 hours for modern epoch. Also we found that the less deep
minimum was assumed to be primary by Uiterdijk's ephemeris.
Further we give the true formulae for minima timings prediction
where the deeper minimum is designated as primary or Min~I. Our
further analysis showed that a less massive component with a lower
temperature is eclipsed at this minimum. This situation is caused
by the current orientation of the orbital ellipse, when at a
deeper minimum the stars are closer to each other and the eclipsed
area of the less brighter star is larger than eclipsed area of the
more brighter star in shallower minimum. More massive and brighter
component is designated "A" and it is eclipsed in secondary
minimum or Min~II. Less massive component with less temperature
named "B" is eclipsed in primary minimum.

The star was observed at the following observatories
(telescope, type of CCD array and photometric system): \\

INASAN Simeiz observatory:\\
 - 60-cm reflector, VersArray~512UV CCD,
$UBVRI$   \\
- 1-m reflector, FLI~PL09000 CCD, FLI~PL18603, $BVRcIc$\\

Crimean observatory of SAI, Nauchny:\\
 - 60-cm reflector, VersArray~1300x1340
CCD, $UBVRI$   \\
  - 60-cm reflector, Ap47p
CCD, $V$   \\
  - 50-cm Maksutov, Pictor~416 CCD, $V$ \\

Tien-Shan high altitude observatory of SAI: \\
- 48-cm reflector, EMI~9863, $WBVR$ \\

Moscow observatory of SAI:\\
 - 70-cm reflector, Ap-7 CCD, $V$  \\

Star\'a Lesn\'a Observatory of the Slovak Academy of Sciences: \\
- 60-cm reflector, Moravian Mono G4-9000 CCD, $V$  \\
- 60-cm reflector, VersArray~512UV CCD, $BVRI$  \\
- 15-cm Maksutov, ST-10XME CCD, $VRcIc$  \\

During minima searching we used every opportunity to obtain
observations of the star, and this explains the wide range of
tools and observatories used. Most of the observations were made
with the 60-cm telescope and VersArray~512UV CCD during 18 nights
in the same instrumental system. Comparison stars
(TYC~5718-588-1=st1 and 2MASS~18584008-1352174=st2, with $V$ =
12.56 and 12.26, respectively, their colour indices are close to
variable) within 3~arcmin in the same field of view as the
variable star were used to determine differential magnitudes.
Normally the variable star differential magnitudes were referenced
to the magnitude of the combined light of both comparison stars
(variable minus comparisons) in each image. Sometimes, when
observing with a CCD of small linear dimensions and a long-focus
telescope, such as 1-m reflector and VersArray~512UV, only
TYC~5718-588-1 was used as a comparison star, as it is only
1.3~arcmin from the variable. All the observations were corrected
for the instrumental systems differences. The magnitudes were
corrected also for nightly variations in the photometric zero
point as we have done previously in similar studies (see,
e.g.,\citet{2010ARep...54..418V}). These corrections found in the
course of Light Curve (LC) solutions reached $\pm 0.01$~mag of the
mean.

A total of 4754 measurements in all photometrical bands were
obtained over 54 nights between 2004 and 2021. All the original
data can be found in a suitable computer form on-line.

We present 24 individual minima timings for V490~Sct, of which
all, without exception, were observed or recalculated (two
photographic timings) in this study and have never been published
elsewhere.

%
\section{Light Curve analysis, colour indices, absolute dimensions}

\begin{figure}
\centerline{\includegraphics[width=\hsize]{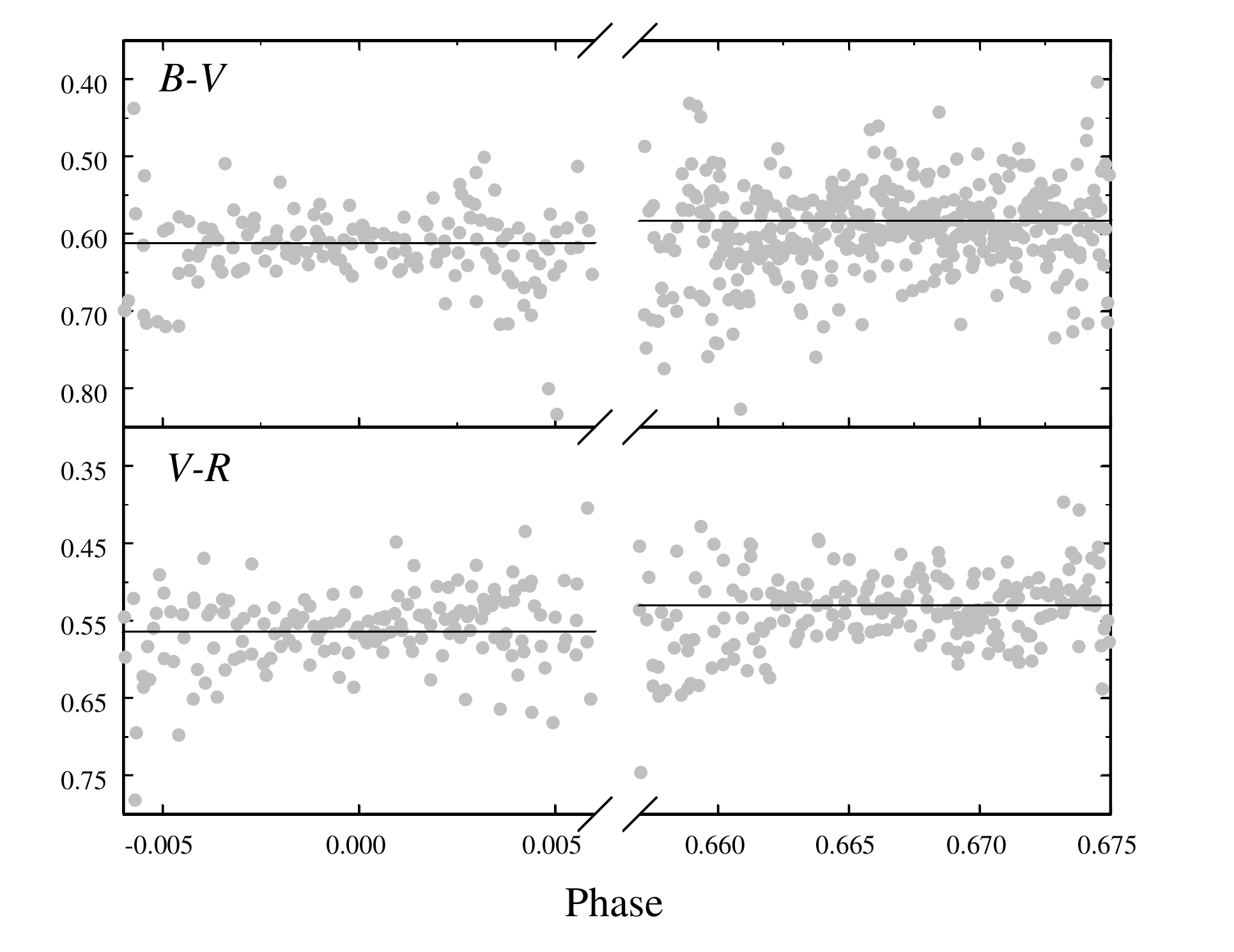}}
\caption{ The colour indices $B-V$ and $V-R$ of the star in both
minima. }
\label{bvvr}
\end{figure}

\begin{figure}
\centerline{
\includegraphics[width=\hsize]{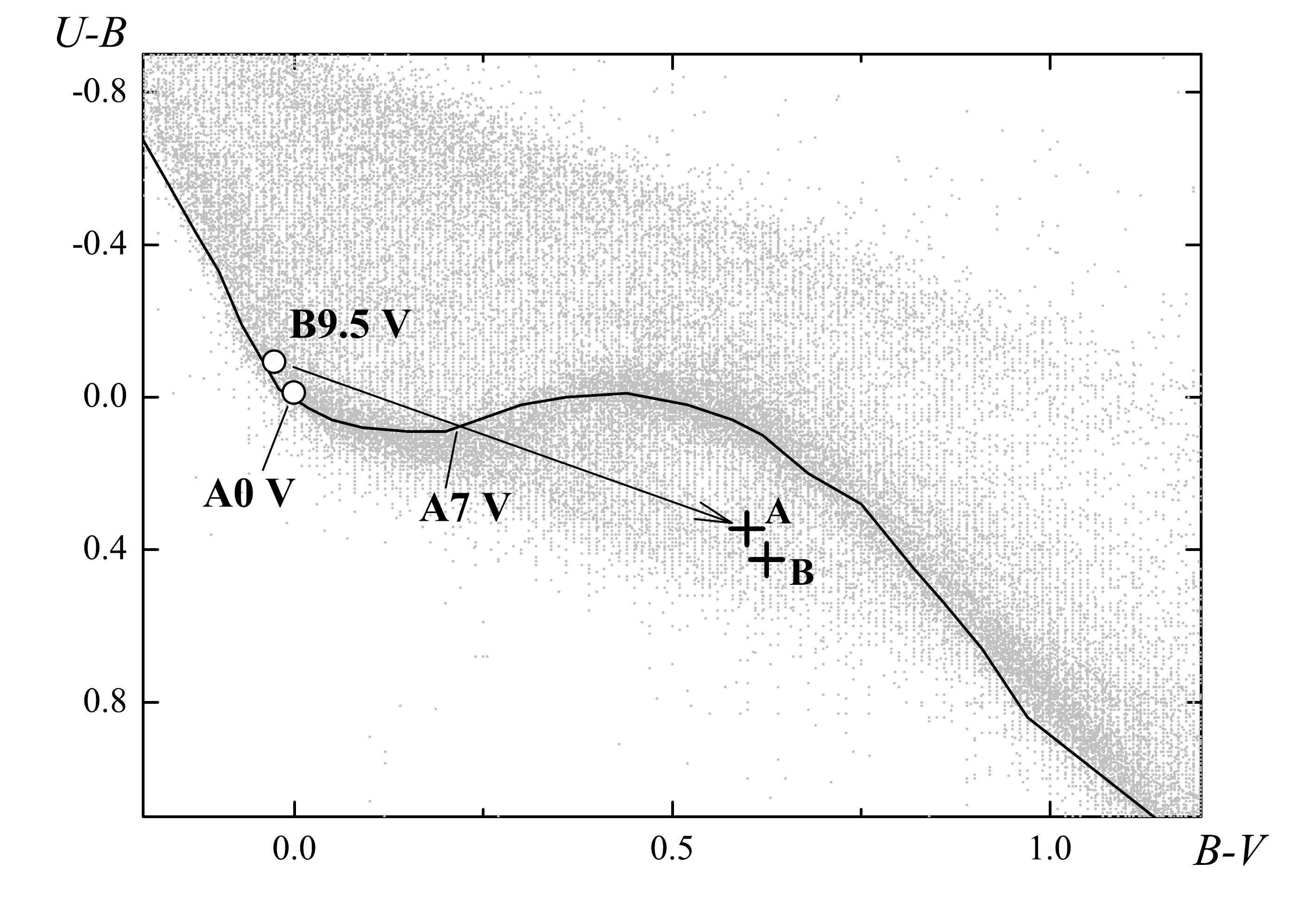}}
\caption{The $(U-B), (B-V)$ diagram. Arrow indicate the direction
of the interstellar reddening. The bold line stands for the
standard luminosity class V sequence, \citet{1992msp..book.....S}.
Cloud of points represents observations in the Johnson $UBV$
system from the \citet{1997yCat.2168....0M} catalogue. Black
crosses mark the observed colour indices of "A" and "B"
components. Empty circles stand for dereddened position of the
colour indices of the components.} \label{ubbv}
\end{figure}

We have derived the magnitudes of the variable and nearby stars
relatively to equatorial standards 109~1082, 110~340,
\citet{1979AJ.....84..627M} with 60-cm cassegrain and
VersArray1340x1300 CCD in Nauchny and to nearby standard star from
SAI catalogue HD171130 \citet{1991TrSht..63....4K} with 48-cm
reflector and $WBVR$ photometer with photomultiplier EMI~9863 at
Tien-Shan observatory. The averaged data of all estimates and
their errors are presented in Table~\ref{magnitudes}.
\begin{table}
\begin{center}
\caption{Magnitudes of V490~Sct in quadratures with comparison
stars.} \label{magnitudes}
\begin{tabular}{cccccc}
\hline \hline
             &  $V$    &  $U-B$  &  $B-V$  &  $V-R$  &  $R-I$   \\
\hline
V490~Sct     & 13.131  &  0.382  &  0.611  &  0.555  &  0.366   \\
             &  0.006  &  0.014  &  0.011  &  0.014  &  0.029   \\
V490~Sct,"A" & 13.828  &  0.344  &  0.599  &  0.540  &  0.362   \\
             &  0.006  &  0.014  &  0.011  &  0.014  &  0.029   \\
V490~Sct,"B" & 13.942  &  0.426  &  0.624  &  0.572  &  0.370   \\
             &  0.006  &  0.014  &  0.011  &  0.014  &  0.029   \\
st1          & 12.560  &  0.327  &  0.447  &  0.454  &  0.320   \\
             &  0.007  &  0.008  &  0.009  &  0.008  &  0.008   \\
st2          & 12.257  &  0.264  &  0.400  &  0.391  &  0.256   \\
             &  0.004  &  0.017  &  0.006  &  0.005  &  0.006   \\
\hline\hline
\end{tabular}
\end{center}
\end{table}
The temperatures of the components can be found as the colour
indices of the light loss in minima. The corresponding
calculations were performed for the most numerous and accurate
$BVR$ observations, see \ref{precision}. The result is presented
in Fig.~\ref{bvvr} which demonstrates that a star with a little
bit less colour indices (what means with higher temperature) is
eclipsed in secondary minimum. Other possibility to get colour
indices is to calculate them from relative luminosities, see
Table~\ref{geometry} obtained in the course of the LCs solutions
in different passbands. Both methods are not completely
independent and gave the same result. To get temperatures of the
stars from their measured colour indices, one should correct them
for interstellar absorption. We can use $(U~-~B),(B~-~V)$
two-colour diagram, see Fig.~\ref{ubbv}. $E(U-B)/E(B-V)~=~0.70$
was accepted for the B9-A0 spectral classes from Table~11 in
\citet{1992msp..book.....S}. The reddening line crosses the fifth
class of luminosity normal sequence in two points - firstly close
to A7 spectral class, the father one near B9.5 -- A0. Both
positions correspond to a different temperatures and masses of the
stars. It is necessary to make a right choice between them.
%
%
Fig.~\ref{ubbv} shows that the straight line connecting the
position of the components in diagram is parallel to the line of
normal colour indices precisely in the B9.5 -- A0 area not in A7,
where it just perpendicular to it. In other words, only B9.5 -- A0
position provides the same interstellar extinction for both
components of the eclipsing system. So for subsequent analysis, we
used the hypothesis of significant interstellar absorption, which
corresponds to the spectral types of B9.5 -- A0 for the
components. The colour indices calculated this way,
\ref{dereddened}, were applied to determine the temperatures with
the help of \citet{1996ApJ...469..355F} calibrations.
\begin{equation}
\begin{array}{c}
\mathrm{"A" component:~}\\
(U-B)_0=-0.094\pm 0.010,~(B-V)_0=-0.027\pm 0.010,\\
(V-R)_0=0.016\pm 0.012\\
\mathrm{"B" component:~}\\
(U-B)_0=-0.012\pm 0.010,~(B-V)_0=-0.002\pm 0.010,\\
(V-R)_0=0.048\pm 0.012
\end{array}
\label{dereddened}
\end{equation}
Let us compare the obtained value of interstellar reddening with
surveys. At the \textit{GAIA} DR2 distance of 1.3 kpc, the
Pan-STARRS 1 3D reddening map \citet{2015ApJ...810...25G}
indicates a reddening of $E(B-V)$~=~0.27~+0.03/-0.01~mag which is
much less than obtained value of $E(B-V)$~=~0.626. Note that the
effect has been already encountered  in the study of young
eclipsing stars with elliptical orbits such as GG~Ori
\citet{2002ARep...46..747V}, V944~Cep \citet{2015ASPC..496..266V},
V2544~Cyg \citet{2017ASPC}, V839~Cep \citet{2019CoSka..49..434V}
and V1103~Cas (unpublished).
These colour indices correspond to $T_A=9960$~K and $T_B=9560$~K
according to \citet{1996ApJ...469..355F} calibration. In the
subsequent analysis the temperature of component "B" had to be
increased by 140K. We used most accurate $B-V$ temperature
calibration, other measured indices $(V-R), (V-I), (V-Rc), (V-Ic)$
do not contradict the obtained values of temperatures.
%
%
\begin{figure}
\centerline{\includegraphics[width=\hsize]{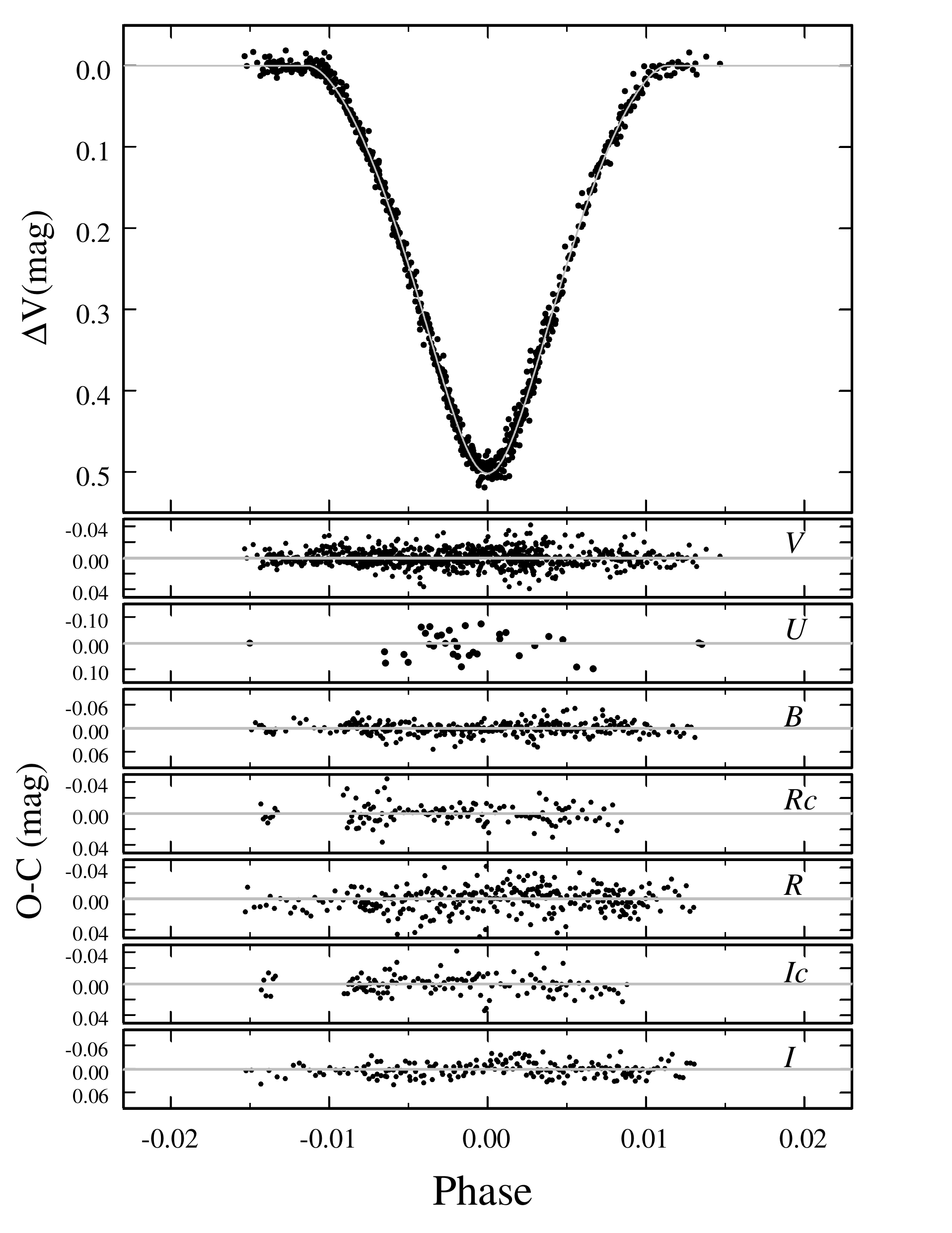}}
\caption{The differential $V$-band observations in Min~I, where
component "B" is eclipsed, upper panel. Out of eclipse brightness
is accepted to be equal zero. Residuals of solutions in different
photometric bands are shown in the lower panels.} \label{Min_I}
\end{figure}
%
\begin{figure}
\centerline{\includegraphics[width=\hsize]{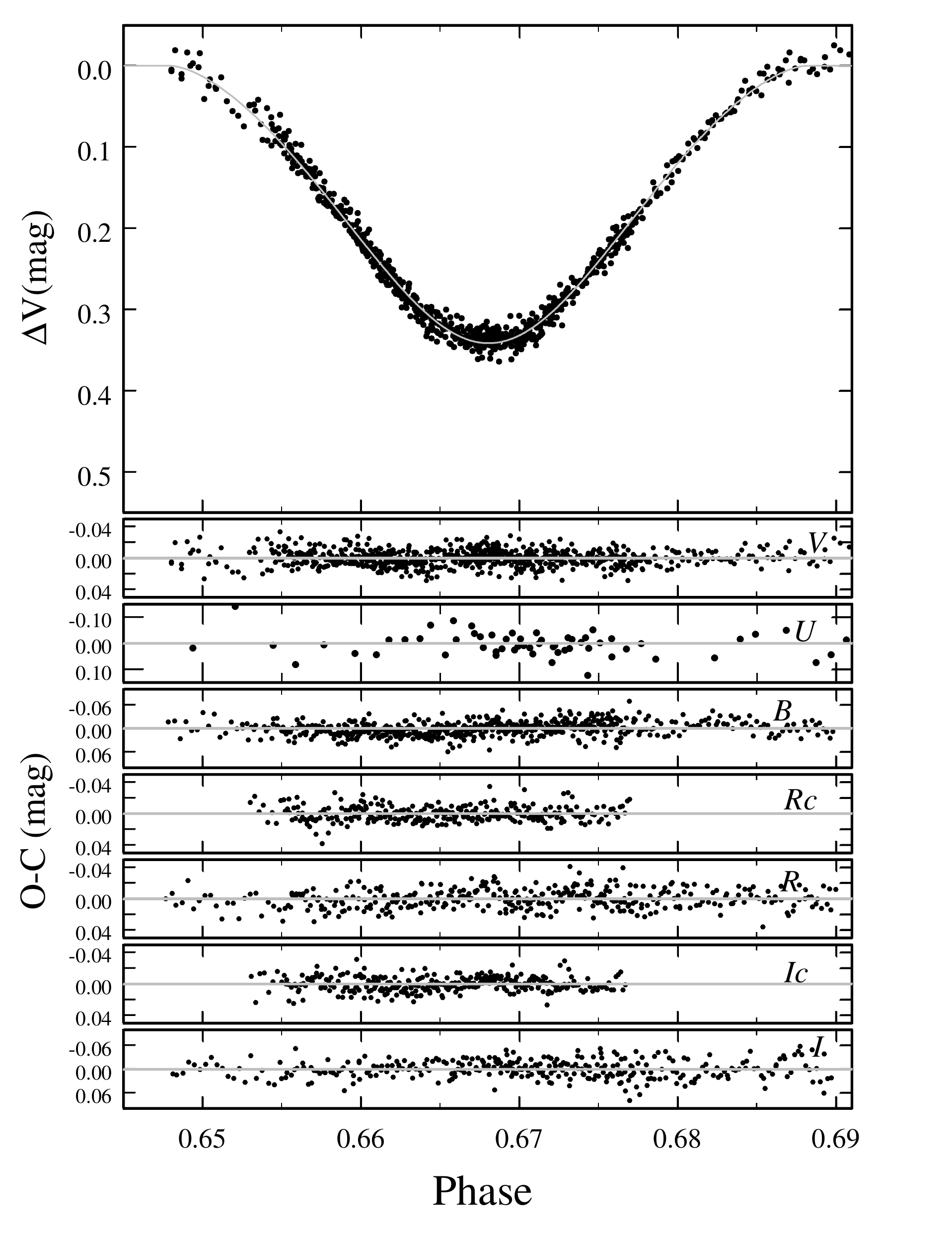}}
\caption{The same as Fig.~\ref{Min_I}, but for Min~II, where
component "A" is eclipsed.} \label{Min_II}
\end{figure}
%
The LCs of the binary show no proximity effects. Therefore, we
used a model of two spherical stars with linear limb-darkening law
moving on an elliptic orbit. We simulated the LCs using our
program based on the algorithm described in
\citet{1984SvA....28..228K}. The limb-darkening coefficients were
fixed according to \citet{1985A&AS...60..471W} for the
temperatures and gravitational acceleration of the components. The
final solution is given in Table~\ref{geometry} and
Table~\ref{absolute}.
%
\begin{table}
\begin{center}
\caption{LCs solution} \label{geometry}
\begin{tabular}{c c}
\hline \hline
Parameter         &  Value     \\
\hline
r$_{A}$           & 0.0520(2)  \\
r$_{B}$           & 0.0507(2)  \\
r$_{A}$+r$_{B}$   & 0.1027(1)  \\
$i^\circ$         & 88.10(3)   \\
$e$               & 0.4008(4)  \\
$\omega^\circ$    & 50.71(9)   \\
L$_{A}U$          & 0.5508(30) \\
L$_{A}B$          & 0.5321(10) \\
L$_{A}V$          & 0.5262(8)  \\
L$_{A}R_c$        & 0.5238(10) \\
L$_{A}R$          & 0.5197(10) \\
L$_{A}I_c$        & 0.5202(20) \\
L$_{A}I$          & 0.5180(20) \\
 \hline\hline
\end{tabular}
\end{center}
\end{table}
%
\begin{figure}
\centerline{\includegraphics[width=\hsize]{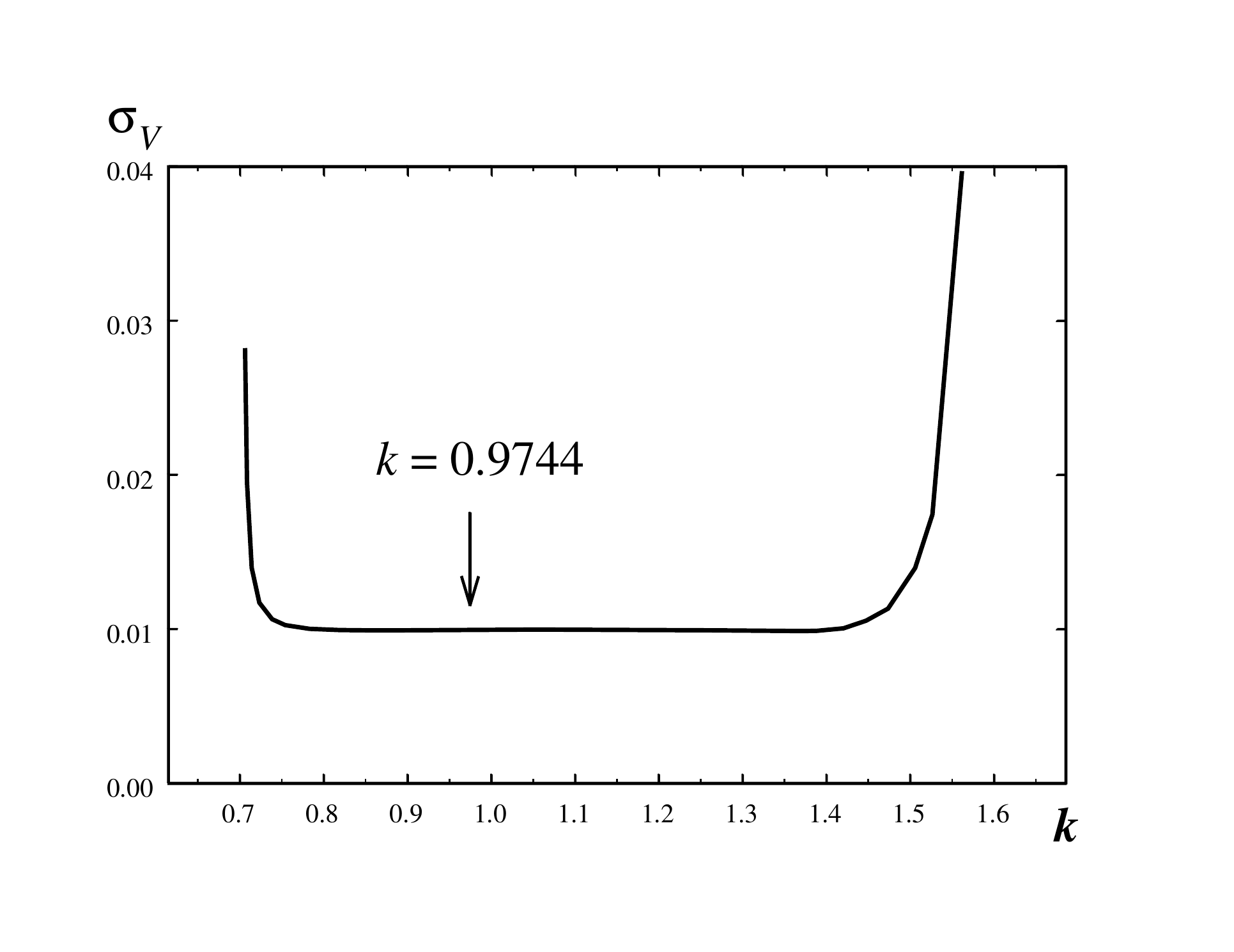}}
\caption{The dependence of the residuals of the LC solutions from
radii ratio $k$. } \label{sigma_k}
\end{figure}

\begin{figure}
\centerline{\includegraphics[width=\hsize]{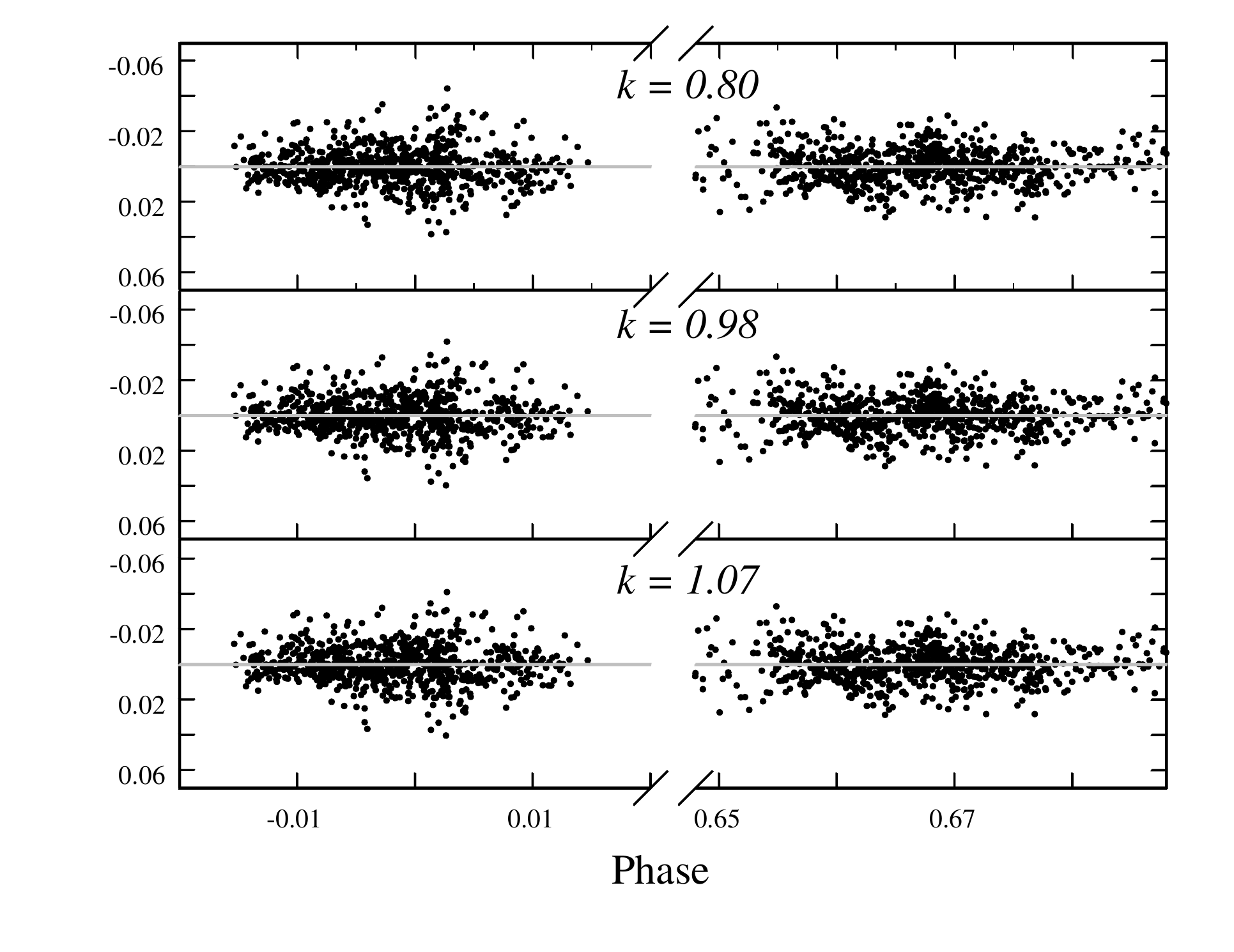}}
\caption{Residuals of our $V$ observations. The values of $k$
fixed as indicated.} \label{diff_k}
\end{figure}

Assuming a normal distribution for the residuals we calculated the
mean errors for an individual point in every spectral band. The
number of points used in calculations is indicated in brackets:
\begin{equation}
\begin{array}{c}
U-0.045~(98),~B-0.0165~(882),~V-0.0104~(1520),\\
R_C-0.0107(413),~R-0.0140~(644),~I_C-0.0105~(432),\\
I-0.021~(502).
\end{array}
\label{precision}
\end{equation}
As it often happens the obtained solutions are insensitive to the
ratio of the radii of the components $k~=r_B/r_A$ for values of
$k$ between 0.7 and 1.3, see Figure~\ref{sigma_k}.
\citet{1982ApJ...254..203P} recommend to make a true choice by
examination of systematic effects of residuals from various
solutions in minima. When we used this recommendation  for our
observations at the minima, we did not find any systematics, see
Figure~\ref{diff_k}.

We point three factors which can explain the failure of the method
in the case. \textbf{1}. V490~Sct has inclination of the orbit
near 88 deg, which implies partial eclipses in which the
dependence of the LC on the ratio of the radii is not as
pronounced as for DI~Her, which Popper used for such analysis.
\textbf{2}. Popper, see Fig.~5 in \citet{1982ApJ...254..203P},
does not consider the photometric zero point in his analysis of
DI~Her, the underestimation of which can also lead to systematic
differences in the residuals. \textbf{3}. Error in darkening
coefficient of the eclipsed component also can produce the same
systematic deviations near conjunction.

\begin{figure}
\centerline{\includegraphics[width=\hsize]{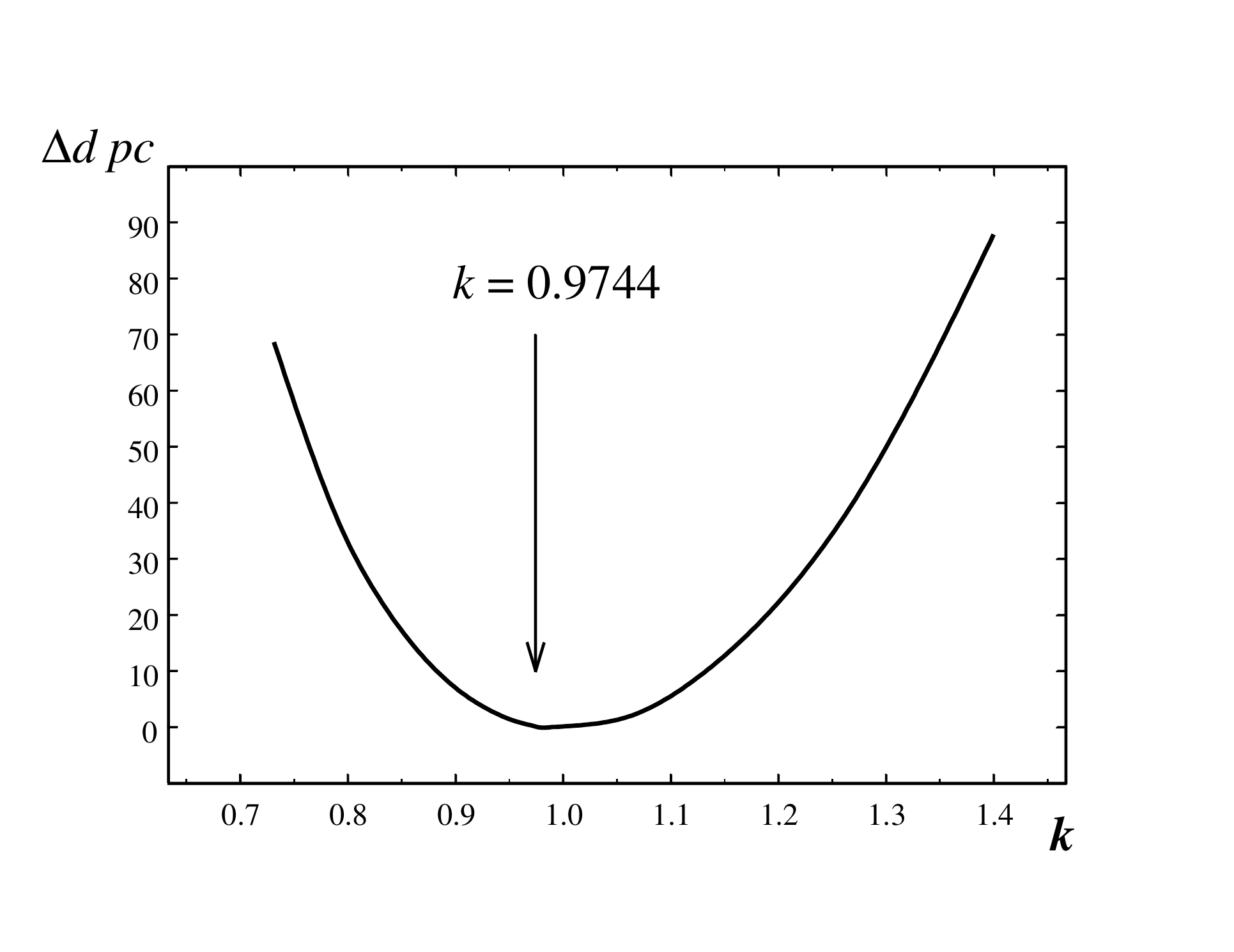}}
\caption{The dependence of components distances from $k$.}
\label{pi_k}
\end{figure}

So, in order to determine the correct value of $k$, we should use
additional information. It's natural to assume that the distances
to both components of the system should be the same. From our
multicolour observations we directly get the temperatures of the
components. Then fixing $k$ in considered range we get a set of
solutions from which we estimate the distances to each component.
The plot in Figure~\ref{pi_k} for most precise $V$-band
observations and temperatures $T_A=9960$~K and $T_B=9700$~K
demonstrates much more pronounced than in Figure~\ref{sigma_k}
minimum at $k$~=~0.9744. Assuming this value we obtain the final
geometrical parameters shown in Table~\ref{geometry}.

The flux ratio in $V$ light, $J_B/J_A$ = 0.9516, corresponds
(\citet{1980ARA&A..18..115P}, Table~1) to a difference in $B-V$ of
0.013 mag or 200K in temperature. Close enough to adopted value of
260K.

%
We estimated the absolute parameters such as semi-major axe, radii
and masses by the non-direct method described in details in
\citet{1985ApJ...299..668K} and \citet{2017CoSka..47...29V}.

Let us estimate the precision of the method.
\begin{table}
\begin{center}
\caption{The absolute parameters derived by the non-direct
method.}
\label{absolute}
\begin{tabular}{ccc}     
\hline \hline
Parameter            & "A" comp.        &  "B" comp.       \\
\hline
$T$ K                &  $9960 \pm60$    &  $9700\pm 80$    \\
B.C.                 & -0.241           &  -0.187          \\
$M$ (M$_\odot$)      &  $2.33\pm0.07$   &  $2.24\pm 0.07$  \\
$q(M_B/M_A$)         & \multicolumn{2}{c}{$0.961\pm0.001$} \\
$R$ (R$_\odot$)      &  $1.91\pm0.04$   & $1.86\pm 0.04$   \\
log $L$ (L$_\odot$)  &  $1.466\pm0.02$  & $1.398\pm 0.02$   \\
log $g$              &  $4.244\pm0.008$ & $4.250\pm 0.008$ \\
$a$ (R$_\odot$)      & \multicolumn{2}{c}{36.65 $\pm $ 0.10} \\
d [pc]               & \multicolumn{2}{c}{$1300\pm40$}     \\
\hline\hline
\end{tabular}
\end{center}
\end{table}
%
The obtained parameters are close to the parameters of other three
eccentric systems derived by a similar method from our own $UBV$
observations which we have calibrated by temperature according to
\citet{1996ApJ...469..355F}. They are V541~Cyg,
\citet{2007A&AT...26..129V}, GG~Ori, \citet{2002ARep...46..747V}
and AS~Cam, \citet{1983Ap&SS..94..115K}. But these systems have
well established masses derived from spectral observations,
correspondingly: \citet{2000AJ....120.3226T},
\citet{2017ApJ...836..177T}, \citet{2011ApJ...734L..29P}.
\begin{table}
\begin{center}
\caption{Masses of the stars.}
\label{masses}
\begin{tabular}{cccc}
\hline \hline
Star           & M obs   & M calc & M obs~-~M calc \\
\hline
  V541~Cyg~"A" &  2.369  &  2.420 & -0.051 \\
  V541~Cyg~"B" &  2.288  &  2.385 & -0.097 \\
    GG~Ori~"A" &  2.336  &  2.301 & +0.035 \\
    GG~Ori~"B" &  2.287  &  2.343 & -0.056 \\
    AS~Cam~"A" &  3.213  &  3.191 & +0.022 \\
    AS~Cam~"B" &  2.323  &  2.305 & +0.018 \\
\hline\hline
\end{tabular}
\end{center}
\end{table}
%
\begin{figure}
\centerline{\includegraphics[width=\hsize]{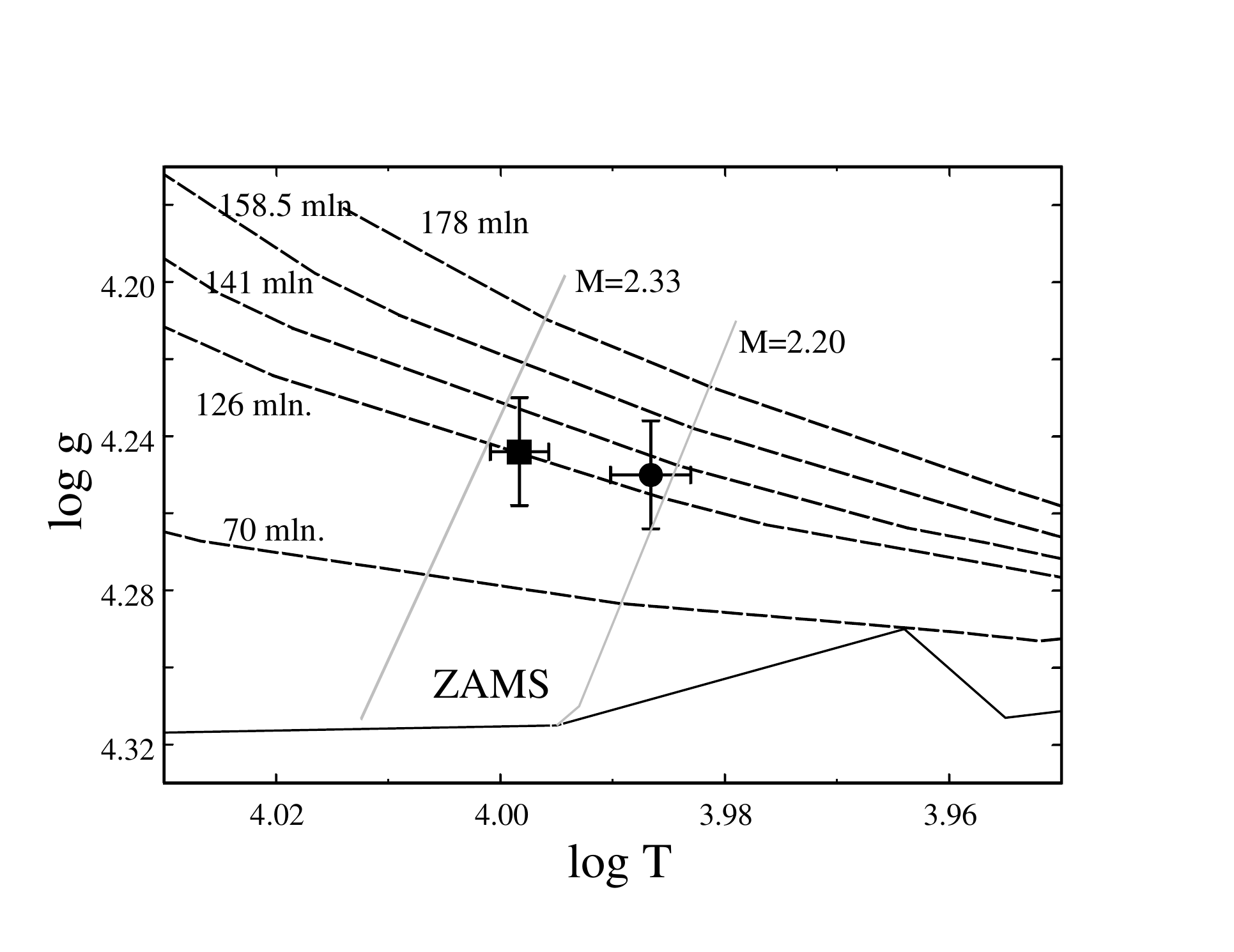}}
\caption{Measurements for V490~Sct in the $log~g$ vs $log~T_{eff}$
diagram compared with evolutionary tracks and isochrones from
\citet{2000A&AS..141..371G} for a Solar metallicity that perfectly
matches the observations. Component "A" - filled square, component
"B" - filled circle. ZAMS is shown with solid line, isochrones for
selected ages are shown with dashed lines. Evolutionary tracks are
presented with solid grey lines.} \label{evolution}
\end{figure}
%
We solved our LCs of these three systems and obtained their
absolute parameters the same way as we did for V490~Sct, see
Table~\ref{masses}. We found that the indirect estimations for
this limited sample are enclosed with probability 68.2\% in $\pm$
0.06 M$_\odot$ interval assuming Student's t-distribution of
errors for this small sample. We conclude that the indirect method
works quite good for the stars similar to V490~Sct. The absolute
parameters of the system are presented in Table~\ref{absolute}.
The derived photometric parallax, $\pi=0.77\pm0.02$ mas, matches
quite well the \textit{GAIA} DR2 value, $\pi=0.76\pm0.04$ mas, of
\citet{2018A&A...616A...9L}. Figure~\ref{evolution} presents the
position of the components of V490~Sct on the evolutionary diagram
$log~g--log~T$ which corresponds to the age of 130 mln. years. The
position of components in $log~L--log~T$ diagram,
Figure~\ref{evolution_L}, demonstrate lack of luminosity in
comparison with theoretical expectations.
%
\begin{figure}
\centerline{\includegraphics[width=\hsize]{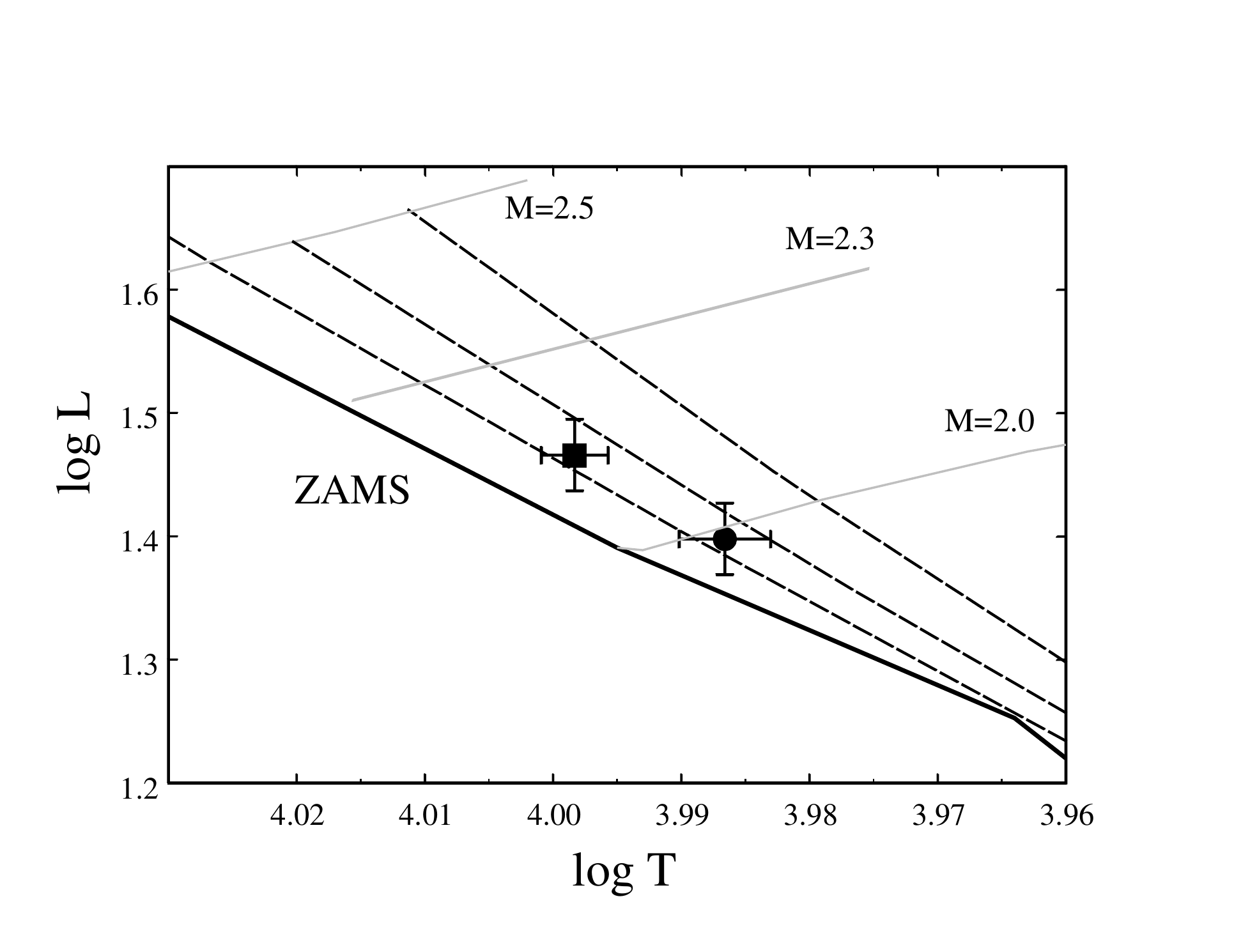}}
\caption{$log~L$ vs $log~T$ diagram. All designations are the same
as in Fig.~\ref{evolution}.} \label{evolution_L}
\end{figure}
%
\section{Apsidal motion}
We've got precise minima timings from our data by fitting the
synthetic LCs to observations obtained during single overnight run
by means of the same program as we used for LCs solution. All the
geometrical parameters were fixed according to their values from
Table~\ref{geometry} except of the specific epoch. In the case of
simultaneous observations in several filters, the minima timings
were weighted and mean values were calculated. The minima timings
are listed in Table~\ref{timings}. Note that no other data exist
at the moment for the star. The same way we obtained mean timings
for photographic observations assuming $B$ band for them. Their
formal weight calculated as 1/$\epsilon^2$ (given in the second
column of Table~\ref{timings}) appeared to be only $10^{-2}$ of
CCD observations weight, but they have sense, as they are 70 years
away from the epoch of our observations. Solving the data from
Table~\ref{timings} by the least squares method separately for the
primary and secondary minima we get the following ephemeris:
\begin{equation}
\begin{array}{c}
{\rm HJD~Min~I~} = 2455073.39094(2)+12.04395915(9)\times E, \\
{\rm HJD~Min~II~} = 2455069.39288(16)+12.0439480(7)\times E. \\
\end{array}
\label{ephemeris}
\end{equation}
The plots which illustrate the residuals given in the 5th column
of Table~\ref{timings} are presented at Figure~\ref{O-C}. The
difference between the periods (\ref{ephemeris}) is evidence of
the rotation of the line of apsides. The rate of apsidal motion
from the periods difference may be found according to formula (6)
from \citet{1989SvA....33...41K}:
%
\begin{figure}
\centerline{\includegraphics[width=\hsize]{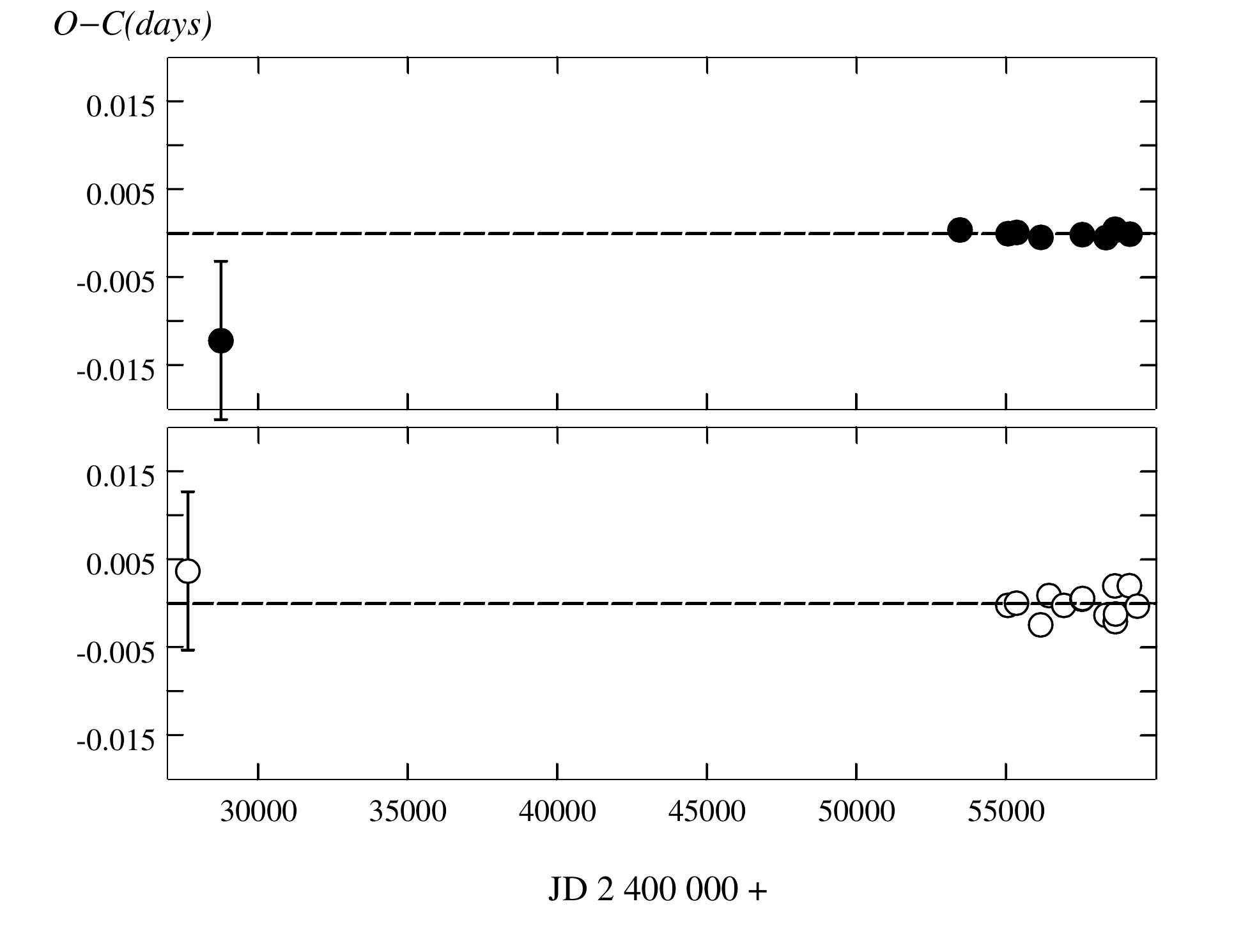}}
\caption{Measured timings of minimum light from
Table~\ref{timings} and ephemeris (\ref{ephemeris}). Filled
circles represent primary minima -- upper panel, empty circles--
secondaries, bottom panel. The errors of CCD timings are less than
the size of the circles.} \label{O-C}
\end{figure}
%
$\dot\omega_{obs}~=~0.0086^\circ$(7)\,$yr^{-1}$,
$U~=~41900(3400)$ years.
Errors in eccentricity and periastron longitude, see
Table~\ref{geometry}, are small and have a little effect on the
resulting error of the measured value. The error in the received
value is ten percent and is caused mainly by errors of the
periods.

We estimate the predicted rate of periastron advance from
classical terms (tidal and rotational distortions) to be
$\dot\omega_{class}~=~0.00219^\circ$(5)\,$yr^{-1}$, where
we have adopted internal structure constants for the two stars of
$log k_{2A}$ = -2.352 and $log k_{2B}$ = -2.355 from the models by
\citet{2004A&A...424..919C} for 130 mln. years age and assuming
that the components are synchronized in periastron. The GR
contribution (e.g., \citet{levi1937astronomical};
\citet{1985ApJ...297..405G}) is calculated to be
$\dot\omega_{rel}~=~0.0103^\circ$(2)\,$yr^{-1}$ which is
4.8 times larger than the classical effect. The total expected
apsidal motion is then
$\dot\omega_{theor}~=~0.0125^\circ$(2)\,$yr^{-1}$. Our
measurement is 32\% less. We can say with confidence that the
observed apsidal rotation is slower than it follows from
synchronism conditions. Deceleration is not as pronounced as in
the case of DI~Her and AS~Cam systems, but is quite noticeable.
The reason for the apparent discrepancy could be the inclination
of axial axes of the system components to the orbital plane,
proposed in the work \citet{1985SvAL...11..224S}.
%
\begin{table}
\begin{center}
\caption{Times of Eclipse for V490~Sct.}
\label{timings}
\begin{tabular}{llcclc}
\hline \hline
HJD            &~~~~$\epsilon$ &  &      & \textit{O - C}  &         \\
(2,400,000+)   & (days)     & Eclipse & Type &   (days)        &  Reference\\
\hline
27657.371      &   0.009   &   2     &  pg  &   +0.0037       &  1      \\
28757.328      &   0.009   &   1     &  pg  & --~0.0122       &  1      \\
53471.5447     &   0.0011  &   1     &  ccd &   +0.0004       &  2      \\
55069.3927     &   0.0007  &   2     &  ccd & --~0.0002       &  2      \\
55073.3909     &   0.0008  &   1     &  ccd & --~0.0000       &  2      \\
55358.4477     &   0.0007  &   2     &  ccd &   +0.0000       &  3      \\
55362.4461     &   0.0002  &   1     &  ccd &   +0.0001       &  3      \\
56165.3897     &   0.0007  &   2     &  ccd & --~0.0024       &  2      \\
56169.3908     &   0.0009  &   1     &  ccd & --~0.0005       &  2      \\
56442.4038     &   0.0001  &   2     &  ccd &   +0.0009       &  3      \\
56936.2046     &   0.0002  &   2     &  ccd & --~0.0002       &  3      \\
57550.4467     &   0.0002  &   2     &  ccd &   +0.0005       &  4      \\
57554.4464     &   0.0001  &   1     &  ccd & --~0.0002       &  4      \\
57562.4907     &   0.0003  &   2     &  ccd &   +0.0006       &  4      \\
58345.3454     &   0.0004  &   2     &  ccd & --~0.0013       &  3      \\
58349.3473     &   0.0001  &   1     &  ccd & --~0.0005       &  3      \\
58646.4474     &   0.0007  &   2     &  ccd &   +0.0020       &  2      \\
58650.4473     &   0.0001  &   1     &  ccd &   +0.0005       &  4      \\
58658.4873     &   0.0005  &   2     &  ccd & --~0.0021       &  3      \\
58662.4911     &   0.0003  &   1     &  ccd &   +0.0003       &  3      \\
58670.5321     &   0.0005  &   2     &  ccd & --~0.0012       &  3      \\
59128.2054     &   0.0004  &   2     &  ccd &   +0.0020       &  3      \\
59144.2490     &   0.0003  &   1     &  ccd & --~0.0001       &  3      \\
59393.1699     &   0.0010  &   2     &  ccd & --~0.0003       &  3      \\
\hline
\end{tabular}
\end{center}
\textbf{Notes.} Columns list the measurement error; the type of
eclipse - 1 primary, 2 - secondary; method of observations;
residuals from the linear fits (\ref{ephemeris}). \\
\textbf{References.} (1) Dr H. van Gent, P.Th.Oosterhoff and
J.Uiterdijk photographic observations reprocessed in present work;
(2) timings from $V$ observations; (3) mean weighted timing from
$BVRI$ observations; (4) mean timing from $BVRcIc$ measurements.
\end{table}

\section{Conclusions}

We obtained reliable parameters for the Algol-type binary
V490~Sct: colour indices, interstellar reddening, masses,
inclination, effective temperatures of the components, radii. The
derived interstellar reddening is anomalously high.

The apsidal motion measured with reliable precision demonstrates
slowness. V490~Sct joined to a very small group of eclipsing
systems with accurately measured apsidal motion in which the
contribution from GR is significant.

The masses of the components computed by a non-direct method are
in agreement with the theoretical diagrams.

We'd like to encourage high-resolution and high signal-to-noise
spectroscopic observations of the system in order to determine the
masses of the components from radial velocity curves.

\begin{acknowledgements}

This study was partly supported by the
scholarship of the Slovak Academic Information Agency(IMV, ASK),
RNF grant 14-12-00146 and RFBR grant 18-502-12025(IMV).

\end{acknowledgements}


\end{document}